\documentclass[pra,twocolumn,showpacs]{revtex4}
\usepackage{epsfig,psfrag}
\usepackage{amsmath,amsfonts}
\usepackage{exscale}
\usepackage{mathrsfs}
\usepackage{times}
\usepackage[T1]{fontenc}
\usepackage{textcomp}
\allowdisplaybreaks

\begin{document}
\renewcommand{\t}[1]{\text{#1}}              
\renewcommand{\d}[1]{\t{d}#1\,}              
\newcommand{\tr}{\mathrm{Tr}}                
\renewcommand{\Re}{\mathrm{Re}\,}            
\renewcommand{\Im}{\mathrm{Im}\,}            
\newcommand{\nfrac}[2]{\genfrac{}{}{0pt}{}{#1}{#2}} 
\renewcommand{\i}{\mathrm{i}}                
\newcommand{\e}[1]{\mathrm{e}^{#1}}          
\title{Simplified approach to double jumps for fluorescing dipole-dipole
  interacting atoms}

\author{Volker Hannstein}                   
\author{Gerhard C.~Hegerfeldt}

\affiliation{Institut f\"ur Theoretische Physik, Universität Göttingen,
  Friedrich-Hund-Platz~1, 37077 G\"ottingen, Germany}

\begin{abstract}
  A simplified scheme for the investigation of cooperative effects in
  the quantum jump statistics of small numbers of fluorescing atoms
  and ions in a trap is
  presented. It allows the analytic treatment of three
  dipole-dipole interacting  four-level systems which model the
  relevant level scheme of Ba$^+$ ions. For the latter, a huge rate of
  double and triple jumps was reported in a former experiment and the
  huge rate was attributed to the dipole-dipole interaction. Our theoretical
  results show that the effect of the dipole-dipole interaction on these rates
  is at most   $5\%$ and that for 
  the parameter values of the experiment there is practically no
  effect. Consequently it seems that the 
dipole-dipole interaction can be ruled out as a possible explanation for
the huge rates reported in the experiment.

\end{abstract}

\maketitle

\section{Introduction}
\label{intro}
The dipole-dipole interaction between atoms and molecules is of
fundamental importance in nature as it gives rise to the all pervading
van der Waals force. In physics, cooperative effects in the radiative
behaviour of atoms due to their mutual dipole-dipole interaction have also
attracted considerable interest in the literature
\cite{refs:FiTaAdBeDaHe}, and they may play a role for possible quantum
computers based on trapped ions or atoms. Atoms 
exhibiting macroscopic light and dark periods in their
fluorescence may provide a sensitive test for such
cooperative effects. Such macroscopic light and dark periods can occur
in a multi-level system if the electron 
is essentially shelved in a metastable state, thereby causing the
photon emission to cease
\cite{ref:DeBeHe}.
Two  such systems accordingly exhibit a dark
period, a bright period of the same intensity as that of a
single system, and a bright period of double intensity. Three
systems exhibit an additional bright period of threefold intensity.
The dipole-dipole interaction may now alter the statistics of
these periods. 
 
In an experiment with two and three Ba$^+$ ions \cite{SaBlNeTo:86,Sa:86} a
large number of double and triple jumps, i.e. jumps by two or 
three intensity steps within a short resolution time, had been
observed, by far exceeding  the number expected for independent atoms.
Theoretically, the quantitative explanation of such large cooperative
effects  for 
distances of the order of ten wave lengths of the strong transition
proved difficult 
\cite{HeNi:88,LeJa:87,LeJa:88,AgLaSo:88,LaLaJa:89,FuGo:92}. 
On the other hand, experiments with different ions showed no observable
cooperative effects \cite{ThBaDhSeWi:92}, in particular none were seen for
Hg$^+$ for a distance of about 15 wave lengths \cite{ItBeWi:88}. 
More recently, effects similar to reference \cite{SaBlNeTo:86} were found in
an experiment with Ca$^+$ ions \cite{BlReSeWe:99}, in contrast to
another, comparable, 
experiment \cite{DoLuBaDoStStStSt:00}. Neither were cooperative effects
found experimentally in an extensive analysis of the quantum jump statistics of
two trapped Sr$^+$ ions \cite{BeRaTa:04}. 
Skornia et al.~\cite{SkoZaAgWeWa:01} recently put forward a new proposal  for observing the dipole-dipole interaction of two V
systems.

\begin{figure}[t,b]
   \psfrag{A1}{$A_1$}
  \psfrag{A2}{$A_2$}
  \psfrag{A3}{$A_3$}
  \psfrag{A4}{\hspace{-0.1cm}$A_4$}
  \psfrag{W}{$W$}
  \psfrag{Omega3}{$\Omega_3$}
  \psfrag{1}{$|1\rangle$}
  \psfrag{2}{$|2\rangle$}
  \psfrag{3}{$|3\rangle$}
  \psfrag{4}{$|4\rangle$}
  \psfrag{6P3/2}{\hspace{-0.3cm}$6\,{}^2\text{P}_{3/2}$}
  \psfrag{6P1/2}{\hspace{-0.3cm}$6\,{}^2\text{P}_{1/2}$}
  \psfrag{6S1/2}{\hspace{-0.3cm}$6\,{}^2\text{S}_{1/2}$}
  \psfrag{5D3/2}{$5\,{}^2\text{D}_{3/2}$}
  \psfrag{5D5/2}{$5\,{}^2\text{D}_{5/2}$}
  \psfrag{(a)}{$\text{(a)}$}
  \psfrag{(b)}{$\text{(b)}$}
  \centering
  \epsfig{file=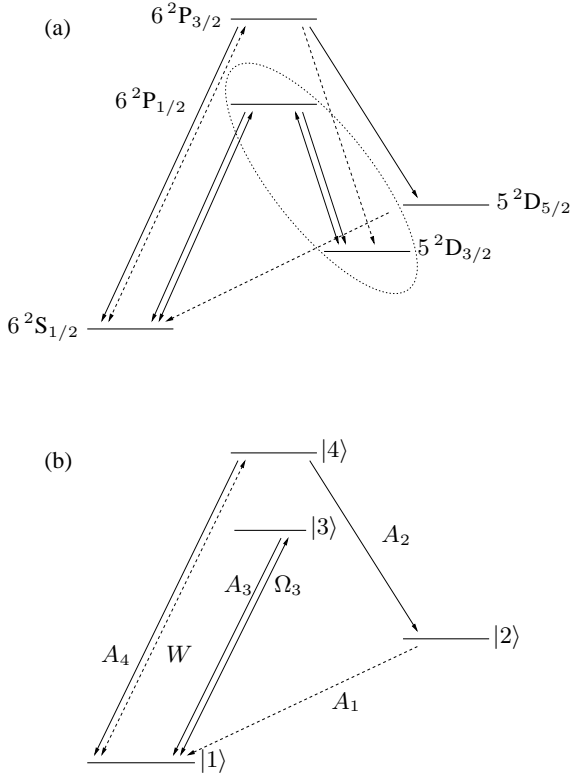, width=6.5cm, height=10cm}
    \caption{\label{fig:5Niveau} (a) Relevant level scheme of 
    $\text{Ba}^+$ \cite{SaBlNeTo:86,Sa:86}. For the effective four
    level system the circled levels are merged to a single level.
    (b) Effective
    four-level system for Ba$^+$. Strong coherent driving of the
    $|1\rangle - |3\rangle$ transition by a laser, weak incoherent
    driving of the $|1\rangle - |4\rangle$ transition by a lamp,
    weak decay of level $|2\rangle$.}
  \hfill
\end{figure}

For two V systems  numerical~\cite{BeHe:99} and
analytical~\cite{AdBeDaHe:01} investigations of the effect of the
dipole-dipole interaction  showed  an increase of up to $30\,\%$
in the double jump rate when compared to independent systems. However, the
systems used in the experimental setups of
references \cite{SaNeBlTo:86,SaBlNeTo:86,ItBeWi:88} were not
 V systems so that a direct comparison between theory and experiment
was not possible. For this reason the present authors extended their
investigation to two other systems \cite{HaHe:03},
namely a D shaped system modeling the Hg$^+$ ions used in reference \cite{ItBeWi:88}
and a four-level system (see Figure \ref{fig:5Niveau}) modeling the
Ba$^+$ ions of references \cite{SaBlNeTo:86,Sa:86}. For two D systems
cooperative effects in the same order of magnitude as for the V
systems were found for ion-distances of a few wavelengths of the
laser-driven transition. For 
larger distances practically no effects where found, in agreement with
the experiments \cite{ItBeWi:88} and with the 
results of reference \cite{SkZaAgWeWa:01b}. In
contrast, only negligible effects for a wide range of ion-distances
were found for two of the four level-systems. Although this result
contradicts the findings of references \cite{SaBlNeTo:86,Sa:86} a direct
quantitative 
comparison with the experiments was not possible since explicit experimental
data were only provided for three Ba$^+$ ions. 
For this reason three of the D and V systems were investigated in
reference \cite{HaHe:04}. In comparison to two of either systems the
cooperative effects found in this case are considerably higher, namely
up to 170$\%$ deviation from the case of independent atoms. However,
since the complexity increases dramatically for higher-level systems, 
this approach could not be applied to three of the four-level
systems which we use to describe the situation of reference \cite{SaBlNeTo:86}. 

In the present paper a simplified approach for the calculation of the
transition rate will be presented with which three four-level systems
can now be treated analytically. This approach is valid for atoms with a
level structure in which the transitions between the different
intensity periods take place incoherently, i.~e.~ via decay or via
incoherent driving. The transition rates for three dipole-interacting
four-level systems will be calculated. Cooperative effects for this
system are found to be less than $5\%$ and negligible for the
experimental parameters of reference \cite{SaBlNeTo:86}. Consequently it
seems that the
dipole-dipole interaction can be ruled out as a possible explanation for
the huge effects measured in the latter experiment.

In section \ref{sec:2} the Bloch equation approach is
recapitulated. On this basis the new method is presented in section
\ref{sec:3} and applied to the four-level systems in \ref{sec:4}. In
section \ref{sec:5} the possibility of a translation of this method to
V system type level structures is discussed.

\section{Bloch equation approach}
\label{sec:2}

\begin{figure}[b]
  \epsfig{file=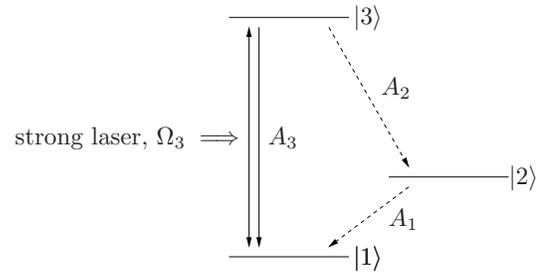,width=8cm}
  \caption{\label{fig:Dsystem} Three-level system in D
      configuration with fast transitions (solid lines) and slow transitions
      (dashed lines).}
\end{figure}

The fluorescence, i.e.~the stochastic sequence of photon emissions, of
a system consisting of a number of atoms with macroscopic bright and
dark periods can be described by a telegraph process. This process is
characterized by the transition rates between the different intensity
periods. In references \cite{AdBeDaHe:01,HaHe:03,HaHe:04} they were
calculated for different model level systems and different numbers of
atoms using a perturbation approach based on the Bloch equation of the
corresponding systems. This approach will be illustrated in the
following by applying it to the simple case of a single three level
system in a D-type configuration as depicted in figure \ref{fig:Dsystem}.

The Bloch equations can be written in the compact form \cite{He:93}
\begin{equation}
\label{bloch}
  \dot{\rho}=-\frac{\text{i}}{\hbar}\left[H_{\text{cond}}\rho-\rho H_{\text{cond}}^{\dagger}\right]+ \mathscr{R}(\rho)
\end{equation}
where $H_{\text{cond}}$ is  the conditional
Hamiltonian of the quantum jump approach \cite{QJ}, for this system
given by 
\begin{equation}
H_\t{cond}=\frac{\hbar}{2\i}\big[ (A_2+A_3)|3\rangle\langle 3|+
A_1|2\rangle\langle 2|\big] +
\frac{\hbar\Omega_3}{2}\big[|1\rangle\langle 3| + |3\rangle\langle 1| \big]
\end{equation}
and $\mathscr{R}(\rho)$ is the the reset state,
\begin{equation}
\mathscr{R}(\rho) = A_1|1\rangle\langle 2|\rho|2\rangle\langle 1| +
A_2|2\rangle\langle 3|\rho|3\rangle\langle 2| +
A_3|1\rangle\langle 3|\rho|3\rangle\langle 1|.
\end{equation}
The Rabi frequency $\Omega_3$ and the Einstein coefficients
$A_1$,$A_2$,$A_3$ are subject to the condition
\begin{equation}
\Omega_3,A_3 \gg A_1,A_2.
\end{equation}
A detuning of the laser has been neglected for simplicity. 
If the small optical parameters $A_1,~A_2$ are neglected the system
splits into independent subspaces. They are given by
\begin{align}
\mathscr{S}_0 & =\{|2\rangle\},& \mathscr{S}_1 = \{|1\rangle,|3\rangle\}
\end{align}
These subspaces $\mathscr{S}_i$ can be associated with the periods of
intensity $I_i$ in the sense that in a period $I_i$ the system is
mostly in the subspace $\mathscr{S}_i$. 
Taking a state $\rho_{0,i}$ in one of the subspaces $\mathscr{S}_i$ at a
time $t_0$ we calculate the state at a time $t_0 + \Delta t$ later
in perturbation theory with respect to the small parameters. The time interval
$\Delta t$ used here should be long compared to the mean time between the
emission of two photons but short compared to the length of the
intensity periods,
\begin{equation}
A_3^{-1},\Omega_3^{-1} \ll \Delta t \ll A_1^{-1},A_2^{-1}~.
\end{equation}
For the calculation the Bloch equation is written in a Liouvillean form,
\begin{equation}
\dot{\rho}=\mathscr{L}\rho =
\{\mathscr{L}_0(A_3,\Omega_3)+\mathscr{L}_1(A_1,A_2)\} \rho~.
\end{equation}
The density matrix at time $t_0+\Delta t$ is then given by \cite{AdBeDaHe:01}
\begin{equation}
\label{eq:rhotDt}
\rho(t_0+\Delta t,\rho_{0,i}) =\rho_{\t{ss},i}+\int_0^{\Delta
  t}\d{\tau}\e{\mathscr{L}_0\tau}\mathscr{L}_1\rho_{\t{ss},i}~, 
\end{equation}
where $\rho_{\text{ss},i}$ is the quasi-steady state in subsystem
$\mathscr{S}_i$, i.e. a steady state of $\mathscr{L}_0$. One can write
\begin{equation}
\mathscr{L}_1\rho_{\t{ss},i}=\sum_{j=0}^1\alpha_{ij}\rho_{\t{ss},j}\Delta
t+ \tilde{\rho},
\end{equation}
with $\tilde{\rho}$ containing the  contributions from the eigenstates
of $\mathscr{L}_0$ for non-zero eigenvalues. This leads to \cite{HaHe:03}
\begin{equation}
\rho(t+\Delta t,\rho_i) = \rho_{\t{ss},i}+\sum_{j=0}^1\alpha_{ij}\rho_{\t{ss},j}\Delta t +(\epsilon-\mathscr{L}_0)^{-1}\tilde{\rho}.
\end{equation}
The last term can be neglected and the coefficient $\alpha_{ij}$ can
therefore be interpreted as transition rate $p_{ij}$ from intensity period $I_i$
to period $I_j$. 
They can be calculated by means of the dual eigenstates for eigenvalue $0$ of
$\mathscr{L}_0$ \cite{HaHe:03}.  
For a single D system the quasi-steady states are given by 
\begin{align}
\rho_{\t{ss},0} & =|2\rangle\langle 2|,\label{ss0} \\ 
\rho_{\t{ss},1} & =
\frac{1}{A_3^2+2\Omega_3^2}\big[(A_3^2+\Omega_3^2)|1\rangle\langle 1|
+\Omega_3^2|3\rangle\langle 3| \nonumber \\ & +\i
A_3\Omega_3|1\rangle\langle 3| - \i A_3\Omega_3|3\rangle\langle 1|
\big]\label{ss1} 
\end{align}
for the dark and the light period respectively. The corresponding dual
states are
\begin{align}
\rho_\t{ss}^0 & = |2\rangle\langle 2|, & \text{and}&& \rho_\t{ss}^1 & =
|1\rangle\langle 1|+|3\rangle\langle 3|
\end{align}
 From  (\ref{ss0}) and (\ref{ss1}) one finds
\begin{subequations}
\label{eq:L1rho1D}
\begin{align}
\label{eq:L1rho1D0}
\mathscr{L}_1\rho_{\t{ss},0} & = - A_1|2\rangle\langle 2|
+A_1|1\rangle\langle 1| \\
\intertext{and}
\label{eq:L1rho1D1}
\mathscr{L}_1\rho_{\t{ss},1} & =
-A_2\frac{\Omega_3^2}{A_3^2+2\Omega_3^2}|3\rangle\langle 3| +
A_2\frac{\Omega_3^2}{A_3^2+2\Omega_3^2}|2\rangle\langle 2| \nonumber
\\ & - \frac{\i 
  A_2}{2}\frac{A_3\Omega_3}{A_3^2+2\Omega_3^2}\left(|1\rangle\langle
  3|-|3\rangle\langle 1|\right).
\end{align}
\end{subequations}
The transition rates are then calculated from
\begin{equation}
\label{alpha}
  p_{ij}=\alpha_{ij}=\tr(\rho_\t{ss}^{j\dagger}\mathscr{L}_1\rho_{\t{ss},i}).
\end{equation}
as
\begin{align}
p_{01} & = \alpha_{01} =  A_1 \\
\intertext{and}
p_{10} & = \alpha_{10}= \frac{A_2\Omega_3^2}{A_3^2+2\Omega_3^2},
\end{align}
in agreement with the direct calculation of the transition rates via
the quantum jump approach.

\section{New simplified approach}
\label{sec:3}

Due to the increased number of levels involved, a calculation of the
transition rates for three dipole-interacting four-level systems
would, although in principal feasible with the methods introduced
above, be even more laborious than for three three-level systems.
It is, however, possible to read off the transition rates without having
to carry out the full calculation. One only needs the quasi-steady states of
the corresponding subsystems. In following this simpler approach will
be presented.

By looking at equation \eqref{eq:L1rho1D} one realizes that the last
step in the calculation, namely the projection onto the dual
eigenstates, although formally more satisfactory, was
actually not necessary in order to gain the final result. The
transition rates are already present as prefactors for some of the density
matrix elements. In fact, $\mathscr{L}_1$ can be interpreted as a
transition operator.
Applying it to some state of the system
yields the density matrix elements which are modified by the weak
decays multiplied by the corresponding decay rates. They are positive for
density matrix element which gain population and negative for those
which loose population due to the decay. In the case in which one
started with $\rho_{\t{ss},0}=|2\rangle\langle 2|$ one therefore has a
term $-A_1|2\rangle\langle 2|$, which accounts for the loss of
population of level $|2\rangle$, and a term $A_1|1\rangle\langle 1|$
for the corresponding gain of population in the ground state. When
starting with $\rho_{\t{ss},1}$ the Einstein coefficient $A_2$ for the
decay from $|3\rangle$ to $|2\rangle$ has an 
additional factor $\frac{\Omega_3^2}{A_3^2+2\Omega_3^2}$ for the
quasi-steady state population of level $|3\rangle$. The last two terms in
equation \eqref{eq:L1rho1D1} are due to the decay of the coherences between
$|1\rangle$ and $|3\rangle$.

From these considerations one is lead to a simple scheme for the
evaluation of the transition rates. First one has to identify the
different independent subspaces for vanishing weak decay rates and
calculate the quasi-steady states in these subspaces as in the above
Bloch equation approach. For a single D system
these are the states $\rho_{\t{ss},0}$ and $\rho_{\t{ss},1}$ for the
subsystems associated with the dark and bright period,
respectively. By looking at the level scheme one can then determine
the possible decay channels between the subsystems.In the present case
this is a decay by $A_2$ from $|3\rangle$ to $|2\rangle$ and a decay
by $A_1$ from $|2\rangle$ to $|1\rangle$. The transition rates are
then given by these decay rates multiplied with the steady state
population of the decaying level.

Physically this is quite intuitive: The transition rates are given by
the corresponding decay rates multiplied by the mean occupation
probabilities of the levels involved. 

\begin{figure}[t,b]
   \psfrag{g}{$|g\rangle$}
  \psfrag{s12}{$|s_{12}\rangle$}
  \psfrag{a12}{$|a_{12}\rangle$}
  \psfrag{e2}{$|e_2\rangle$}
  \psfrag{s23}{$|s_{23}\rangle$}
  \psfrag{a23}{$|a_{23}\rangle$}
  \psfrag{e3}{$|e_3\rangle$}
  \psfrag{s13}{$|s_{13}\rangle$}
  \psfrag{a13}{$|a_{13}\rangle$}
  \psfrag{(a)}{$\text{(a)}$}
  \psfrag{(b)}{$\text{(b)}$}
  \centering
  \epsfig{file=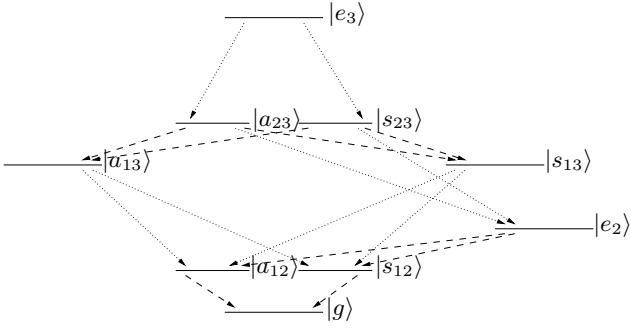, width=8cm}
    \caption{\label{fig:2Dniveau} {Level configuration of two D systems in the
      Dicke basis. Transitions with rate $A_2 \pm \Re C_2$
      (dotted arrows) and transitions with rate $A_1 \pm
      \Re C_1$ (dashed arrows). Fast transitions (with $A_3 \pm \Re
      C_3$) and line shifts due to detuning and to $\Im C_i$ are omitted.}} 
\end{figure}

The question is now if this approach can be extended to more
complicated systems, especially to dipole-interacting D systems and to
the four-level system for the description of Ba$^+$. This is indeed
possible. For two dipole-interacting D systems for example the
possible decays can be read off  Figure \ref{fig:2Dniveau} which
shows the level scheme in the Dicke basis given by
\begin{gather}
|g\rangle = |1\rangle|1\rangle,\quad |e_2\rangle = |2\rangle|2\rangle, \quad
  |e_3\rangle = |3\rangle|3\rangle
\nonumber \\
  |s_{ij}\rangle = \frac{1}{\sqrt{2}}\big(|i\rangle|j\rangle +
  |j\rangle|i\rangle\big), \\
  |a_{ij}\rangle =
  \frac{1}{\sqrt{2}}\big(|i\rangle|j\rangle -
  |j\rangle|i\rangle\big)~. \nonumber 
\label{Dicke} 
\end{gather}
The easiest case is the transition rate $p_{01}$ for a transition from
a dark period to a period of intensity $I_1$. Here the relevant transitions
 are from $|e_2\rangle$ to $|s_{12}\rangle$ and $|a_{12}\rangle$. The
corresponding decay rates are $A_1+\Re C_1$ and $A_1-\Re C_1$,
respectively, with the dipole-dipole coupling parameters $C_i$ given
explicitly in reference \cite{HaHe:03}. The
quasi-steady state population of $|e_2\rangle$ is unity, 
so the transition rate is $p_{01}=2A_1$, in agreement with the result
of reference \cite{HaHe:03}. The other transition rates are a bit more
complicated. For $p_{10}$ one has to take into account the decays from
$|s_{23}\rangle$ and $|a_{23}\rangle$ to $|e_2\rangle$, for $p_{12}$
the decays from $|s_{23}\rangle$ and $|a_{23}\rangle$ to
$|s_{13}\rangle$ and from $|s_{12}\rangle$ and $|a_{12}\rangle$ to
$|g\rangle$, and for $p_{21}$ the decays from $|e_3\rangle$ to
$|s_{23}\rangle$ and $|a_{23}\rangle$ and from $|s_{13}\rangle$ and
$|a_{13}\rangle$ to $|s_{12}\rangle$ and $|a_{12}\rangle$. 
Multiplying for each decay the decay rate by the steady state
population of the initial level and adding up the different
contributions then yields the same results for the transition rates
as obtained by the Bloch equation approach in reference
\cite{HaHe:03}. The same is also true for
three dipole-interacting D systems \cite{HaHe:04}.

\section{Three dipole-interacting four-level systems}
\label{sec:4}

An application of the simplified method to the four-level system describing
Ba$^+$ is also possible. As depicted in Figure \ref{fig:5Niveau} the
transition from a bright to a dark period is a two step process for
this system, first an excitation to level $|4\rangle$ by incoherent
light with the rate $W$ and then a decay to level $|2\rangle$ with the
Einstein coefficient $A_2$. Instead of a single Einstein coefficient
one therefore has to use the product of the incoherent transition rate
$W$ with the branching ratio $A_2/(A_2+A_4)$ for a decay from state
$|4\rangle$ to state $|2\rangle$ for this transition. Then everything
works as in the case of the D systems and one confirms the results for
a single four-level system already known from the Bloch equation
approach \cite{HaHe:03}. 

Consequently it is also possible to obtain the transition rates for
three four-level systems which would be rather involved to do with the
Bloch equation approach. The Bloch equations can be written in the
compact form
\begin{eqnarray}
  \label{Bloch4N}
  \dot{\rho} & = & -\frac{\i}{\hbar} \Big[ H_\t{cond} \rho - \rho
  H_\t{cond}^\dagger\Big] +
  \mathscr{R}_W(\rho) + \mathscr{R}(\rho)\\
  & \equiv & \left\{\mathscr{L}_0 +
    \mathscr{L}_1(A_1,W)\right\}\rho, \nonumber
\end{eqnarray}
where $\mathscr{R}_W(\rho)$ describes the incoherent driving as
in reference \cite{HePl:93} and is given by
\begin{equation}
\mathscr{R}_W(\rho)= W\sum_{i=1}^3 \big(S_{i4}^+\rho S_{i4}^- +S_{i4}^-\rho S_{i4}^+\big)~,
\end{equation}
with 
\begin{align*}
  S_{i1}^+& =|2\rangle_i {}_i\langle 1|, & S_{i2}^+& =|4\rangle_i
  {}_i\langle 2|, & S_{i3}^+ & =|3\rangle_i {}_i\langle 1|  \\ S_{i4}^+ & =|4\rangle_i {}_i\langle
  1|, & & \text{and}
  & S_{ij}^- & =S_{ij}^{+ \dagger}~.
\end{align*}
The conditional Hamiltonian, without detuning, and the reset state in
this case are given by
\begin{align}
\label{Hcond}
  H_{\text{cond}} & = 
  \sum_{i=1}^3\sum_{j=1}^4 \frac{\hbar}{2\text{i}}A_j S_{ij}^+S_{ij}^- + \sum_{i=1}^3 \frac{\hbar}{2} \left[
    \Omega_3S_{i3}^-  + \mbox{h.c.} \right] \nonumber \\* 
   & + \sum_{\nfrac{k,l=1}{k<l}}^3\sum_{j=1}^4
  \frac{\hbar}{2\text{i}} C_{kl}^{(j)}\left(S_{kj}^+ S_{lj}^- +
    S_{lj}^+S_{kj}^-\right)
\end{align}
and
\begin{align}
  \label{reset}
  \mathcal{R}(\rho) & = \sum_{i=1}^3\sum_{j=1}^4 A_j S_{ij}^-\rho
  S_{ij}^+ \nonumber \\
  & + \sum_{\nfrac{k,l=1}{k<l}}^3\sum_{j=1}^4
  \text{Re}\,C_{kl}^{(j)}\left(S_{kj}^-\rho S_{lj}^+ + S_{lj}^-\rho
    S_{kj}^+\right), 
\end{align}
where
\begin{eqnarray}
  C_{kl}^{(j)} & = &
  \frac{3A_j}{2}\e{\i a_{kl}^{(j)}}\left[\frac{1}{\i a_{kl}^{(j)}}(1
    -\cos^2 \theta_{kl}) \right. \\ && \left. \hspace{1cm} {} 
    + \left(\frac{1}{a_{kl}^{(j)2}}-\frac{1}{\i
        a_{kl}^{(j)3}}\right)(1-3\cos^2 \theta_{kl})\right] \nonumber 
\end{eqnarray}
is the coupling parameter which describes the dipole-dipole interaction
between atom $k$ and atom $l$ for the transition connected with the
Einstein coefficient $A_j$, with $\theta_{kl}$ being the angle between
the dipole moments and the line connecting the atoms. The dimensionless
parameter $a_{kl}^{(j)}=2\pi r_{kl}/\lambda_j$ is given by the
inter-atomic distance $r_{kl}$ multiplied by the wave number
$2\pi/\lambda_j$ of this transition. In order to get a maximal effect
of the dipole-dipole interaction we assume as in 
\cite{HaHe:04} that the atoms form an equilateral triangle
(i.e.~$r_{kl}=r$) and that $\theta_{kl}= \pi/2$. Then $C_{kl}^{(j)}$
becomes the $C_j$ of reference \cite{HaHe:03}.

The quasi-steady states are already known from the
calculations for three three-level systems. As in reference \cite{HaHe:04}
one can use a symmetrized basis analoguous to the Dicke basis for two
atoms. This leads to the states
\begin{subequations}
\label{dicke3}
\begin{align}
  |s_{ijk}\rangle & = 
  \frac{1}{\sqrt{6}}\big(|i\rangle|j\rangle|k\rangle
    +|j\rangle|k\rangle|i\rangle +|k\rangle|i\rangle|j\rangle 
  \nonumber \\* & \hspace{0.7cm} 
  {}+ |i\rangle|k\rangle|j\rangle + |j\rangle|i\rangle|k\rangle +
    |k\rangle|j\rangle|i\rangle \big), \\
  |a_{ijk}\rangle & = 
  \frac{1}{\sqrt{6}}\big(|i\rangle|j\rangle|k\rangle +|j\rangle|k\rangle|i\rangle
    +|k\rangle|i\rangle|j\rangle \nonumber \\* & \hspace{0.7cm}
  {}- |i\rangle|k\rangle|j\rangle -
    |j\rangle|i\rangle|k\rangle - |k\rangle|j\rangle|i\rangle \big), \\
  |b_{ijk}\rangle & = \frac{1}{\sqrt{12}}\big(2
    |i\rangle|j\rangle|k\rangle -|j\rangle|k\rangle|i\rangle
    -|k\rangle|i\rangle|j\rangle \nonumber \\* & \hspace{0.9cm} 
  {}+ 2|i\rangle|k\rangle|j\rangle -
    |j\rangle|i\rangle|k\rangle - |k\rangle|j\rangle|i\rangle \big), \\
  |c_{ijk}\rangle & =
  \frac{1}{2}\big(|j\rangle|k\rangle|i\rangle -
    |k\rangle|i\rangle|j\rangle \nonumber \\* & \hspace{0.7cm}
  {}- |j\rangle|i\rangle|k\rangle +
    |k\rangle|j\rangle|i\rangle \big), \\*
  |d_{ijk}\rangle & = \frac{1}{\sqrt{12}}\big(2|i\rangle|j\rangle|k\rangle
    -|j\rangle|k\rangle|i\rangle -|k\rangle|i\rangle|j\rangle
  \nonumber \\* & \hspace{0.9cm} 
  {}-2|i\rangle|k\rangle|j\rangle +|j\rangle|i\rangle|k\rangle
    +|k\rangle|j\rangle|i\rangle \big), \\ 
  |e_{ijk}\rangle & = \frac{1}{2}\big(|j\rangle|k\rangle|i\rangle -
    |k\rangle|i\rangle|j\rangle \nonumber \\ & \hspace{0.7cm} 
  {}+ |j\rangle|i\rangle|k\rangle - |k\rangle|j\rangle|i\rangle \big) ~,
\end{align}
\end{subequations}
$i<j<k;\; i,j,k=1,\ldots,4$, in the case where all three atoms are in
different states. For the
remaining states one gets for $i,j=1,\ldots,4$, $i \neq j$,
\begin{subequations}
  \begin{align}
    |s_{ijj}\rangle & =
    \frac{1}{\sqrt{3}}\big(|i\rangle|j\rangle|j\rangle +
    |j\rangle|j\rangle|i\rangle + 
      |j\rangle|i\rangle|j\rangle \big) \\
    |b_{ijj}\rangle & =
    \frac{1}{\sqrt{6}}\big(2|i\rangle|j\rangle|j\rangle -
      |j\rangle|j\rangle|i\rangle - |j\rangle|i\rangle|j\rangle \big) \\
    |c_{ijj}\rangle & = \frac{1}{\sqrt{2}}\big(|j\rangle|j\rangle|i\rangle -
      |j\rangle|i\rangle|j\rangle\big) 
  \end{align}
\end{subequations}
if two atoms are in the same state and 
\begin{align}
  |g\rangle & = |1\rangle|1\rangle|1\rangle, & |e_i\rangle & =
  |i\rangle|i\rangle|i\rangle & \mbox{for} \quad i& =2,3,4 
\end{align}
if all three atoms are in the same state. 
The quasi-steady states for intensity periods $I_0$ to $I_2$ are, by
symmetry, given by 
\begin{subequations}
\begin{align}
\rho_{\t{ss},0}& =|e_2\rangle\langle e_2| \\
\rho_{\t{ss},1}& =
\frac{1}{3}\big\{\rho_\t{ss}^\t{1D}\otimes|2\rangle_2 {}_2\langle
  2|\otimes|2\rangle_3 {}_3\langle 2| 
  \\ & \hspace{-0.5cm} + |2\rangle_1 {}_1\langle
  2|\otimes\rho_\t{ss}^\t{1D}\otimes|2\rangle_3 {}_3\langle 2| + |2\rangle_1 {}_1\langle
  2|\otimes|2\rangle_2 {}_2\langle 2|\otimes\rho_\t{ss}^\t{1D}\big\} \nonumber\\
\rho_{\t{ss},2}& =
\frac{1}{3}\sum_{i=1}^3\rho_{\t{ss},2}^\t{2D}\otimes|2\rangle_i
{}_i\langle 2|, 
\end{align}
\end{subequations}
where $\rho_\t{ss}^\t{1D}$ is the quasi-steady state of one D system in the
$\{|1\rangle,|3\rangle\}$ subspace and $\rho_{\t{ss},2}^\t{2D}$ is the
quasi-steady state in the subspace corresponding to double intensity of two
D systems. The state $\rho_{\t{ss},3}$ is rather
complicated. Therefore only the populations of the relevant levels 
will be given, i.e.
\begin{subequations}
\begin{gather}
\langle g|\rho_{\t{ss},3}|g\rangle =
\frac{1}{N}\Big[\big\{(A_3^2+\Omega_3^2) \big[(A_3^2+\Omega_3^2)^2 +
3A_3^2B\big] \nonumber \\ 
 +2A_3\big[|C_3|^2|A_3+C_3|^2 +B^2\big]\big\}\Big],\\
\langle s_{113}|\rho_{\t{ss},3}|s_{113}\rangle =
\frac{\Omega_3^2}{N}\big[(A_3^2+\Omega_3^2)(3A_3^2+\Omega_3^2)+ 3A_3^2B\big] \\
\langle b_{113}|\rho_{\t{ss},3}|b_{113}\rangle =\langle
c_{113}|\rho_{\t{ss},3}|c_{113}\rangle =
\frac{\Omega_3^4}{N}(A_3^2+\Omega_3^2) \\
\langle s_{133}|\rho_{\t{ss},3}|s_{133}\rangle =
\frac{\Omega_3^4}{N}(3A_3^2+\Omega_3^2) \\
\langle e_3|\rho_{\t{ss},3}|e_3\rangle =\langle
b_{133}|\rho_{\t{ss},3}|b_{133}\rangle = 
\langle c_{133}|\rho_{\t{ss},3}|c_{133}\rangle =
\frac{\Omega_3^6}{N}
\end{gather}
with
\begin{eqnarray*}
N&=&\big\{(A_3^2+2\Omega_3^2) \big[(A_3^2+2\Omega_3^2)^2 +
3A_3^2B\big] 
\\ & &+2A_3\big[|C_3|^2|A_3+C_3|^2 +B^2\big]\big\}
\end{eqnarray*}
and
\[
B = |C_3|^2+2A_3\Re C_3.
\]
\end{subequations}

Now the procedure is the same as described in the previous section for
two D systems and one obtains
\begin{subequations}
\label{eq:pij}
\begin{align}
\label{eq:uppij}
p_{01}& = 3A_1 & p_{12}& = 2A_1 & p_{23} & = A_1 
\end{align}
and 
\begin{align}
  p_{10} & = \frac{A_2W(A_3^2+\Omega_3^2)}{(A_2+A_4)[A_3^2 +
    2\Omega_3^2]} \\
  p_{21} & = \frac{2A_2W}{A_2+A_4}\Bigg[\frac{A_3^2+\Omega_3^2}{A_3^2
    + 2\Omega_3^2} + 2\,\Re C_3\frac{A_3^3\Omega_3^2}{[A_3^2 +
    2\Omega_3^2 ]^3}\Bigg] 
  \\
  p_{32} & = \frac{3A_2W}{A_2+A_4}\Bigg[\frac{A_3^2+\Omega_3^2}{A_3^2
    + 2\Omega_3^2} + 4\,\Re C_3\frac{A_3^3\Omega_3^2}{[A_3^2 +
    2\Omega_3^2 ]^3}\Bigg]
\end{align}
\end{subequations}
as transition rates up to first order in $C_3$. The exact results
including detuning are given in the appendix. The approximations to
first order in $C_3$ have the same structure as for three
dipole-interacting three-level systems given in reference
\cite{HaHe:04}. Basically this means an increase of cooperative
effects by a factor of two compared to two atoms. In terms of these
transition rates the double and triple jump rate, i.e.~the rate of two
or three subsequent jumps within a short time window $T_W$, are then
given by \cite{HaHe:03} 
\begin{figure}[tb]
  \psfrag{nDJ [10^-3/s]}{$\quad n_\t{DJ} [10^{-3}\t{s}^{-1}]$}
  \psfrag{r [l3]}{$r [\lambda_3]$}
  \epsfig{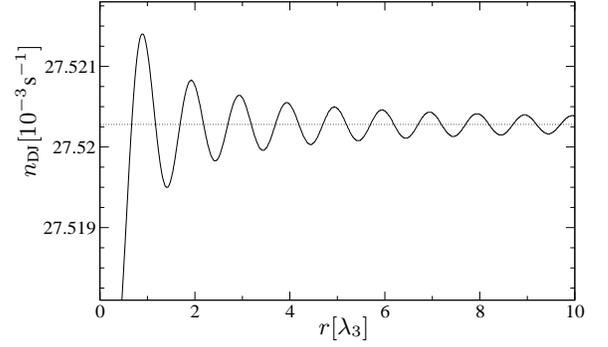}
  \caption{\label{fig:34nivnDj}  Double jump rate $n_\t{DJ}$ and
    for three dipole-interacting four-level systems with experimental
    parameter values of reference \cite{SaNeBlTo:86}. Dotted line:
    independent systems. Cooperative effects are less than
    1\textperthousand~ for distances larger than $\lambda_3$.}
\end{figure}
\begin{equation}
n_\t{DJ}=2\frac{p_{01}p_{21}p_{32}(p_{01}+p_{12})}{p_{21}p_{32}(p_{01}+p_{10})+
  p_{01}p_{12}(p_{23}+p_{32})}T_W
\end{equation}
and 
\begin{equation}
n_\t{TJ} =
2\frac{p_{01}p_{10}p_{12}p_{21}p_{23}p_{32}}{p_{21}p_{32}(p_{01}+p_{10})+
  p_{01}p_{12}(p_{23}+p_{32})}T_W^2~.
\end{equation}

\begin{figure}[tb]
  \psfrag{nTJ [10^-6/s]}{$\quad n_\t{TJ} [10^{-6}\t{s}^{-1}]$}
  \psfrag{r [l3]}{$r [\lambda_3]$}
  \epsfig{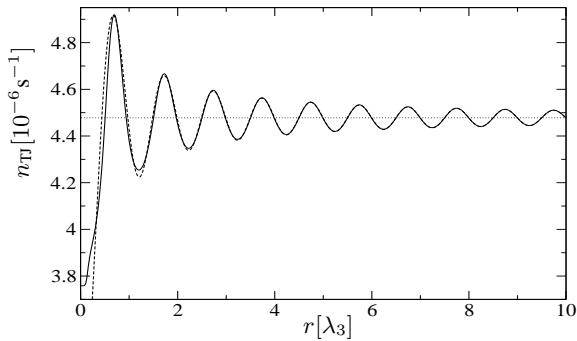}
  \caption{\label{fig:34nivnTj}  Triple
    jump rate $n_\t{TJ}$ for three dipole-interacting four-level 
    systems. Parameter values as in Fig.~\ref{fig:34nivnDj} except for
    $\Delta_3=0$ and $\Omega_3=\frac{1}{2}\sqrt{\!\sqrt{5}-1}A_3$ for maximal
    effects. Dotted line: independent systems, dashed line: up to
    first order. Cooperative effects are less than
    5\% for distances larger than $\lambda_3$.}
\end{figure}

In Fig. \ref{fig:34nivnDj} a plot of $n_\t{DJ}$ for
the experimental parameter values of reference \cite{SaNeBlTo:86} is
shown. The effects of the dipole-dipole interaction are negligibly
small in particular for experimental distances of about ten times the
wavelength $\lambda_3$ of the strong transition. Without detuning
$\Delta_3$, maximal cooperative effects are obtained for
$\Omega_3=\frac{1}{2}\sqrt{\!\sqrt{5}-1}A_3$. This case is shown in
Fig.~\ref{fig:34nivnTj} for the triple jump rate $n_\t{TJ}$.  For
inter-atomic distances  larger than one wavelength $\lambda_3$ of the
strong transition cooperative effects are less than $5\%$ and again
rapidly decreasing for larger distances. For non-zero detuning the
maximally achievable effects have about the same value. Also one has
to bear in mind that, as in reference \cite{HaHe:04}, this result has
to be seen as an upper limit for all possible configurations in the
trap. Large cooperative effects, i.e.~enhancements of the double and
triple jump rate by several orders of magnitude, can therefore not be
explained by the dipole-dipole interaction.   
Furthermore one sees that the first order results of equation
\eqref{eq:pij} are a very good approximation to the exact transition rates
given in the appendix. 
 
\section{V system and similar level schemes}
\label{sec:5}

From the previous results the question arises whether the method presented
here is also applicable to level systems like the V system,
i.e. systems in which the transition between different light and dark
periods results from a coherent excitation. It turns out that for
these systems the situation is much more complicated. For a single V
system for example, $\mathscr{L}_1(\Omega_2)$ contains coherences
between the ground state $|1\rangle$ and the metastable state
$|2\rangle$. Therefore $\mathscr{L}_1\rho_i$ has no component in the
subspace of eigenstates of $\mathscr{L}_0$ for eigenvalue zero and the
state at time 
$t_0+\Delta t$ in the Bloch equation approach is thus given by
\cite{AdBeDaHe:01} 
\begin{equation}
\rho(t_0+\Delta t, \rho_{0,i})= \rho_{\t{ss},i}+
(\epsilon-\mathscr{L}_0)^{-1}\mathscr{L}_1\rho_{\t{ss},i}. 
\end{equation}
An explicit evaluation of this expression for a single V system
starting with $\rho_1$ not only leads to terms proportional to the
quasi-steady state population of the ground state  but also to terms
proportional to the quasi-steady state coherence between ground state and
excited state. 

The situation gets even more involved for dipole-interacting V
systems. Here the term $(\epsilon-\mathscr{L}_0)^{-1}$ gives rise to
additional factors which depend in a very complicated way on
$C_3$. This is in contrast to the D and the four-level system, for
which the $C_3$ dependence in the transition rates is solely due to
the $C_3$ dependence of the quasi-steady states. 
The physical reason for this is that the efficiency of the laser driving is
influenced by the dipole interaction, for example via additional
detunings. Therefore the mechanism of jumps in the light intensity
based on laser driven transitions is much more
complex than for jumps based on  spontaneous decay and incoherent
driving so that the method outlined above is applicable only in the
latter case.

\section{Conclusions}
\label{sec:6}

In this paper we have presented a simplified approach for the calculation
of the transition rates between periods of different intensity of a
system of dipole-dipole interacting atoms which show macroscopic
quantum jumps in their fluorescence. This method works for atoms with
level configurations in which the transition between the different
intensity periods is based on incoherent processes. Results previously
obtained  with other methods are recovered by the new approach.

In addition, the  new method has allowed the calculation of the
transition rates for three 
interacting four-level systems modeling the the relevant level structure of
Ba$^+$ ions. This allows a direct comparison with the experiment of
reference \cite{SaNeBlTo:86}. This experiment reported an
enhancement of the double and triple jump rate by several orders of
magnitude  and this was explained through cooperative effects due to
the dipole-dipole interaction between the ions. With the present
results it is seen that this cannot be the explanation
for the reported enhancement. Cooperative effects can indeed be found for this
system but they are much smaller, namely only maximal
$5\%$ of the 
values for independent atoms. For the parameter values of the
experiment they are practically absent.

\begin{appendix}
\section{Exact transition rates including detuning}

As mentioned above the transition rates between the different
intensity periods can be calculated exactly in $C_3$ and with
inclusion of a possible detuning of the laser $\Delta_3$ with respect
to the corresponding atomic transition. The result for the downward
rates is
\begin{subequations}
\begin{align}
  p_{10} = & \frac{A_2W(A_3^2+\Omega_3^2+4\Delta_3^2)}{(A_2+A_4)[A_3^2 +
    2\Omega_3^2+4\Delta_3^2]} 
\end{align}
\begin{widetext}
\begin{align}
  p_{21} = & \frac{2A_2W}{A_2+A_4}\frac{(A_3^2+\Omega_3^2+4\Delta_3^2)(A_3^2 +
    2\Omega_3^2 +4\Delta_3^2)+(A_3^2+4\Delta_3^2)(|C_3|^2+2A_3\Re C_3-4\Delta_3\Im C_3)}
  {(A_3^2+2\Omega_3^2+4\Delta_3^2)^2+(A_3^2+4\Delta_3^2)(|C_3|^2+2A_3\Re
    C_3-4\Delta_3\Im C_3)}\nonumber  \\
  = &
  \frac{2A_2W}{A_2+A_4}\Bigg[\frac{A_3^2+\Omega_3^2+4\Delta_3^2}{A_3^2 +
    2\Omega_3^2+4\Delta_3^2} + 2\,\Re C_3\frac{A_3\Omega_3^2(A_3^2+4\Delta_3^2)}{[A_3^2 +
    2\Omega_3^2 +4\Delta_3^2]^3} -
  4\,\Im C_3\frac{\Delta_3\Omega_3^2(A_3^2+4\Delta_3^2)}{[A_3^2 +
    2\Omega_3^2 +4\Delta_3^2]^3}\Bigg] + \mathcal{O}(C_3^2). \\
p_{32} = &
\frac{3A_2W}{A_2+A_4}\nonumber \\
&\frac{(A_3^2+\Omega_3^2+4\Delta_3^2)[(A_3^2 +
  2\Omega_3^2 +4\Delta_3^2)^2 \!+ \!3(A_3^2+4\Delta_3^2)B]+\! 2(A_3^2+4\Delta_3^2)
  [|C_3|^2|A_3-2\i\Delta_3+C_3|^2+B(\Omega_3^2+B)]}
  {(A_3^2 + 2\Omega_3^2+4\Delta_3^2)\left[(A_3^2
      +2\Omega_3^2+4\Delta_3^2)^2 \!+\! 3(A_3^2+4\Delta_3^2)B\right] +
    2(A_3^2+4\Delta_3^2)\left[|C_3|^2|A_3\!-\!2\i\Delta_3+C_3|^2
      +B^2\right]} \nonumber \\ 
= & \frac{3A_2W}{A_2+A_4}\Bigg[\frac{A_3^2+\Omega_3^2+4\Delta_3^2}{A_3^2 + 2\Omega_3^2+4\Delta_3^2} +
  4\,\Re C_3\frac{A_3\Omega_3^2(A_3^2+4\Delta_3^2)}{[A_3^2 +
    2\Omega_3^2 +4\Delta_3^2]^3} - 8\,\Im
  C_3\frac{\Delta_3\Omega_3^2(A_3^2+4\Delta_3^2)}{[A_3^2
    +2\Omega_3^2+4\Delta_3^2]^3} \Bigg] + \mathcal{O}(C_3^2) 
\end{align}
\end{widetext}
\end{subequations}
with $B = |C_3|^2+2A_3\Re C_3-4\Delta_3\Im C_3$. The upward rates of
equation \eqref{eq:uppij} are already the exact results since they are
independent of $C_3$ and $\Delta_3$.

\end{appendix}

\end{document}